\documentclass[aps,superscriptaddress,twocolumn]{revtex4}

\usepackage{latexsym}
\usepackage{amssymb}
\usepackage{amsmath}
\usepackage{graphicx}
\usepackage{bm}
\usepackage{epsf}
\usepackage{epstopdf}
\usepackage[colorlinks,
            citecolor=blue,
            anchorcolor=red,
            menucolor=red,
            linkcolor=red,
            filecolor=red,
            runcolor=red,
            urlcolor=blue,
            frenchlinks=red]{hyperref}

\newcommand{\fm}{\mathrm{fm}}
\usepackage[usenames]{color}
\usepackage[normalem]{ulem}  %

\begin{document}

\newcommand{\YSU}{Key Laboratory for Microstructural Material Physics of Hebei Province, School of Science, Yanshan University, Qinhuangdao, 066004, China}
\newcommand{\QDU}{Science School, Qingdao University of Technology, Qingdao 266000, China}
\newcommand{\RCQDU}{The Research Center of Theoretical Physics, Qingdao University of Technology, Qingdao 266033, China}
\newcommand{\PeU}{School of Science, China University of Petroleum (East China), Qingdao 266580, China}

\title{Symmetry energy effect on hot nuclear matter and proto-neutron stars}

\author{Xuhao Wu}~\email{wuhaobird@gmail.com}
\affiliation{\YSU}
\author{Peng-Cheng Chu}~\email{kyois@126.com}
\affiliation{\QDU}
\affiliation{\RCQDU}
\author{Min Ju}~\email{jumin@upc.edu.cn}
\affiliation{\PeU}
\author{He Liu}~\email{liuhe@qut.edu.cn}
\affiliation{\QDU}
\affiliation{\RCQDU}

\begin{abstract}
%====================================================================================================
We examine the effects of symmetry energy on proto-neutron stars (PNSs) by employing equation of state (EOS) described by the relativistic mean-field (RMF) model.
The thermal properties of dense matter and the bulk properties of PNSs are investigated under the assumptions of isothermy, isentropy, and fixed lepton fractions.
The polytropic index is calculated at finite temperature, 
revealing a negative correlation with the PNS maximum mass that the EOS can support.
The properties of PNSs during the heating and cooling stages along their evolution line are explored under different combinations of lepton fraction and entropy.
We investigate the correlation between the symmetry energy slope $L$ and the properties of PNSs. As $L$ increases, the PNS radius also increases; however, this effect diminishes with a growing lepton fraction for the isentropic case.
These results indicate that the nuclear symmetry energy and its density dependence play crucial roles in determining the properties of PNSs and their evolution stages.
\end{abstract}

\maketitle

%%%%%%%%%%%%%%%%%%%%%%%%%%%%%%%%%%%%%%%%%%%%%%%%%%%%%%%%%%%%%%%%%%%%%
\section{Introduction}
% finite temperature
The equation of state (EOS) relating pressure to energy density (or mass density) plays a key role in determining the global structure of cold neutron stars (NSs) older than a few minutes,
which have temperatures below 1 MeV~\cite{Suwa2014,Oertel2017}.
In contrast, when considering studies of thermal dynamics,
such as core-collapse supernovae (CCSNe)~\cite{Shen1998,Yamasaki2006,Constantinou2014,Mabanta2018,Burrows2024}, 
proto-neutron star (PNS) or proto-quark star evolution~\cite{Prakash2001,Shao2011,Chu2017,Raduta2020,Sumiyoshi2022,Chu2024}, and binary neutron star (BNS) mergers~\cite{Shibata2011,Ruiz2020}, 
it requires an EOS input with finite temperature ($T$) and
a fixed lepton fraction ($Y_L$) that is not in weak-equilibrium~\cite{Oertel2017,Li2024}. 
At this point, pressure becomes a function of three thermodynamic parameters, density, temperature and lepton fraction.
In the case of supernovae simulations and PNSs, 
the typical temperature of nuclear matter can range from a fraction of 1 MeV to several tens of MeV.
During CCSNe, electron-capture reactions lead to significant neutronization,
resulting in a decrease in $Y_L$ (which equals $Y_p$, the proton fraction).
In the subsequent cooling of the PNS, $\beta$ equilibrium is approached with neutrinos being free.
Prior to this, the presence of trapped neutrinos limits the minimum value of $Y_L$, 
which is usually less than that of symmetric nuclear matter, i.e. 
$Y_L<0.5$.
To describe the PNS forms following CCSNe of massive stars,
it is necessary to consider baryon number densities up to several
times the nuclear saturation density, temperatures up to 100 MeV, and lepton fractions up to $Y_L\sim0.4$.
%For BNS mergers, the temperature can rise above 100 MeV, and $Y_L$ in ejecta could reach $\sim0.6$.

% Symmetry energy
It is well known that symmetry energy and its slope play an crucial role in the structure of cold NSs~\cite{Ducoin2010,Cavagnoli2011,Providencia2013,Oertel2017}.
During CCSNe and PNS evolution, the symmetry of nuclear matter changes differently compared to cold NS matter.
The temperature and density dependence of the nuclear symmetry free energy, using microscopic two- and three-body nuclear potentials, is investigated in~\cite{Wellenhofer2015}.
The sensitivity of symmetry energy at twice the saturation density as a control parameter to properties of hot
and dense matter is investigated in~\cite{Nakazato2019}.
The influence of symmetry energy on hot nuclear matter for supernova simulations over time is examined in~\cite{Sumiyoshi2019}.
The effect of symmetry energy becomes one of the most significant factors affecting the EOS of hot dense nuclear matter and the bulk properties of PNSs.

% aim
In this work, we investigate the symmetry energy effects on hot dense nuclear matter and the cooling stage of PNS at finite temperature, entropy and lepton fraction. 
We apply the TM1e model~\cite{Bao2014,Shen2020} and the original TM1 model~\cite{Sugahara1994,Shen1998,Shen2011},
both of which have identical properties for symmetric nuclear matter but exhibit different behaviors regarding symmetry energy.
The TM1e model has a symmetry energy slope ($L$) of $40$ MeV, significantly smaller than the $L = 110.8$ MeV found in the original TM1 model. 
By considering the dependence on temperature and symmetry energy, 
we investigate the thermodynamic properties of PNS matter and its bulk characteristics.
Our results show that the symmetry energy dependence can significantly influence the EOS of hot nuclear matter, 
the polytropic index curve, the bulk properties of PNS, and the evolution stages of PNS.

%The YL effect on neutron star and quark star are very different

This paper is organized as follows. 
In Sec.~\ref{sec:model}, we derive the theoretical formalism for PNS matter at finite temperature within the relativistic mean-field (RMF) model.
In Sec.~\ref{sec:results}, we present our calculation results regarding the properties of PNS matter and the PNS itself.
Finally, in Sec.~\ref{sec:summary}, we offer the summary and conclusions.

\section{Hot equation of state}
\label{sec:model}
To describe dense nuclear matter at finite temperatures,
we employ the RMF theory,
which conforms well with terrestrial experimental data and the saturation properties.
The Lagrangian reads~\cite{Meng2006,Bao2014,Bao2016,Shen2020}
\begin{eqnarray}
\label{eq:RMF}
\mathcal{L}_{\rm{RMF}} & = & \sum_{i=p,n}\bar{\psi}_i
\left\{i\gamma_{\mu}\partial^{\mu}-\left(M+g_{\sigma}\sigma\right) \right.\notag \\
&& \left. -\gamma_{\mu} \left[g_{\omega}\omega^{\mu} +\frac{g_{\rho}}{2}\tau_a\rho^{a\mu}
\right]\right\}\psi_i  \notag \\
&& +\frac{1}{2}\partial_{\mu}\sigma\partial^{\mu}\sigma -\frac{1}{2}%
m^2_{\sigma}\sigma^2-\frac{1}{3}g_{2}\sigma^{3} -\frac{1}{4}g_{3}\sigma^{4}
\notag \\
&& -\frac{1}{4}W_{\mu\nu}W^{\mu\nu} +\frac{1}{2}m^2_{\omega}\omega_{\mu}%
\omega^{\mu} +\frac{1}{4}c_{3}\left(\omega_{\mu}\omega^{\mu}\right)^2  \notag
\\
&& -\frac{1}{4}R^a_{\mu\nu}R^{a\mu\nu} +\frac{1}{2}m^2_{\rho}\rho^a_{\mu}%
\rho^{a\mu} \notag \\
&& +\Lambda_{\rm{v}} \left(g_{\omega}^2
\omega_{\mu}\omega^{\mu}\right)
\left(g_{\rho}^2\rho^a_{\mu}\rho^{a\mu}\right) \notag\\
&& +\sum_{l=e,\mu}\bar{\psi}_{l}
  \left( i\gamma_{\mu }\partial^{\mu }-m_{l}\right)\psi_l,
\end{eqnarray}
where $W^{\mu \nu }$ and $R^{\alpha \mu \nu }$ are the  antisymmetric field tensors associated with $\omega^\mu$ and $\rho^{\alpha\mu}$, respectively.

The effective mass of nucleon is expressed as
\begin{eqnarray}
 M{_N^ * }=M+g_{\sigma}\sigma.
\end{eqnarray}
The parameters of mass and coupling constants are listed in Table~\ref{tab:rmf}.
%%%%%%%%%%%%%%%
\begin{table*}[htp]
	\caption{Parameters for the TM1 and TM1e models~\cite{Shen1998,Bao2014}.
		The masses are given in MeV. }
	\begin{center}
		\setlength{\tabcolsep}{2.1mm}{
			\begin{tabular}{lcccccccccccc}
		\hline\hline
				Model   &$M$  &$m_{\sigma}$  &$m_\omega$  &$m_\rho$  &$g_\sigma$  &$g_\omega$
				&$g_\rho$ &$g_{2}$ (fm$^{-1}$) &$g_{3}$ & $c_3$ & $\Lambda_v$ \\
        \hline
				TM1     &938.000  &511.198  &783.000  &770.000  &10.0289  &12.6139  &9.2644
				&7.2325   &0.6183  &71.3075 &0   \\
        \hline
				TM1e    &938.000  &511.198  &783.000  &770.000  &10.0289  &12.6139  &13.9714
				&7.2325   &0.6183  &71.3075 &0.0429   \\
		\hline\hline
		\end{tabular}}
		\label{tab:rmf}
	\end{center}
\end{table*}
%%%%%%%%%%%%%%%

%By solving these equations Eq.~(\ref{sigma})-Eq.~(\ref{electric neutrality}) self-consistently, we can derive the physical quantities we need to calculate the EOS.

The baryon number density can be written as
\begin{eqnarray}
 {n_i} &=&  \frac{1}{{{\pi ^2}}}\int {{k^2}\left( {f_i^k - f_{\bar i }^k} \right)dk} ,
\end{eqnarray}
with the Fermi-Dirac distribution functions
\begin{eqnarray}
  f_{i}^{k} &=& \frac{1}{1+\exp\left[\left(\sqrt{k^2+{M_{i}^{*}}^2}-\nu_{i}\right)
        /T\right]},\\
 f_{\bar{i}}^{k} &=& \frac{1}{1+\exp\left[\left(\sqrt{k^2+{M_{i}^{*}}^2}+\nu_{i}\right)/T\right]} ,
\end{eqnarray} 
$i$ and $\bar{i}$ denote the particle and antiparticle, respectively.
The chemical potential $\mu_i$ reads, 
\begin{eqnarray}
 \mu_{p}       &=& \nu_p      +g_{\omega}\omega +\frac{1}{2}g_{\rho}\rho, \\
 \mu_{n}       &=& \nu_n      +g_{\omega}\omega -\frac{1}{2}g_{\rho}\rho,
\end{eqnarray}
in which $\nu_i$ is the effective chemical potential.
For isothermal or isentropic nuclear matter, the conditions of chemical equilibrium and charge neutrality are satisfied, 
which can be written as
\begin{eqnarray}
 {\mu _n} &=& {\mu _p} + {\mu _e},\  {\mu _\mu } = {\mu _e},\\
\label {chemical potential}
 {n_p} &=& {n_e} + {n_\mu }
\label {electric neutrality}.
\end{eqnarray}
In neutrino trapped matter, the lepton fraction is fixed as $Y_L=\textit{constant}$ together with the chemical equilibrium,
\begin{eqnarray}
{\mu_n}={\mu_p}+{\mu_e}+\mu_{\bar{\nu}_e}.
\end{eqnarray}
In contrast, for neutrino free matter, the lepton fraction $Y_L$ varies with density under $\beta$-equilibrium.

With given number density $n_b$, temperature $T$ (or entropy $s$) and fixed lepton fraction $Y_L$, the expressions for energy density, the pressure and the entropy density can be obtained as
\begin{eqnarray}
 \varepsilon  &=&   \sum\limits_{i = n,p} {\frac{1}{{{\pi ^2}}}\int  {{{\left( {{k^2} + M{{_N^ * }^2}} \right)}^{1/2}}{k^2}\left( {f_i^k + f_{\bar i}^k} \right)dk} }  \nonumber\\
&&   + \frac{1}{2}m^2_{\sigma}{\sigma}^2+\frac{1}{3}{g_2}{\sigma}^3
     +\frac{1}{4}{g_3}{\sigma}^4   \nonumber  \\
&&   + \frac{1}{2}m^2_{\omega}{\omega}^2 + \frac{3}{4}{c_3}{\omega}^4    \nonumber  \\
&&   + \frac{1}{2}m^2_{\rho}{\rho}^2
     + 3{\Lambda}_{\textrm{v}}\left(g^2_{\omega}{\omega}^2\right)
     \left(g^2_{\rho}{\rho}^2\right) \nonumber  \\
  && +\sum\limits_{l = e,\mu } {\frac{1}{{{\pi ^2}}}\int  {{{\left( {{k^2} + m_l^2} \right)}^{1/2}}{k^2}\left( {f_l^k + f_{\bar l}^k} \right)dk} } ,
\label{eq:eoshe}    
\end{eqnarray}
\begin{eqnarray}
 P &=&    \frac{1}{3}\sum\limits_{i = n,p} {\frac{1}{{{\pi ^2}}}\int  {\frac{{{k^4}}}{{{{\left( {{k^2} + M{{_N^ * }^2}} \right)}^{1/2}}}}\left( {f_i^k + f_{\bar i}^k} \right)dk} }   \nonumber\\
&&   - \frac{1}{2}m^2_{\sigma}{\sigma}^2-\frac{1}{3}{g_2}{\sigma}^3-\frac{1}{4}{g_3}{\sigma}^4   \nonumber \\
&&   + \frac{1}{2}m^2_{\omega}{\omega}^2 +\frac{1}{4}{c_3}{\omega}^4      \nonumber \\
&&  + \frac{1}{2}m^2_{\rho}{\rho}^2 
    +\Lambda_{\textrm{v}}\left(g^2_{\omega}{\omega}^2\right)
      \left(g^2_{\rho}{\rho}^2\right) \nonumber \\
&&  +\frac{1}{3}\sum\limits_{l = e,\mu } {\frac{1}{{{\pi ^2}}}\int  {\frac{{{k^4}}}{{{{\left( {{k^2} + m_l^2} \right)}^{1/2}}}}\left( {f_l^k + f_{\bar l}^k} \right)dk} }  ,  
\end{eqnarray}
and
\begin{eqnarray}
s &=& \sum\limits_{i = n,p} {\frac{1}{{{\pi ^2}}}\int {dk\left[ { - f_i^k\ln } \right.f_i^k - \left( {1 - f_i^k} \right)\ln \left( {1 - f_i^k} \right)} }  \nonumber\\
&& \left. { - f_{\bar i }^k\ln f_{\bar i }^k - \left( {1 - f_{\bar i }^k} \right)\ln \left( {1 - f_{\bar i }^k} \right)} \right]   \nonumber\\
&&+\sum\limits_{l =e,\mu} {\frac{1}{{{\pi ^2}}}\int {dk\left[ { - f_l^k\ln } \right.f_l^k - \left( {1 - f_l^k} \right)\ln \left( {1 - f_l^k} \right)} }  \nonumber\\
&& \left. { - f_{\bar l }^k\ln f_{\bar l }^k - \left( {1 - f_{\bar l }^k} \right)\ln \left( {1 - f_{\bar l }^k} \right)} \right] .
\label{eq:eoshs}
\end{eqnarray}
The free energy density reads
\begin{eqnarray}
f=\varepsilon-Ts.
\end{eqnarray}

%$\gamma={d({\mathrm{ln}P})}/{d({\mathrm{ln}\epsilon})}$
A newly born PNS contains trapped neutrinos that
it is standard to consider the leptons numbers as fixed.
Along the evolution line from a PNS to a cold NS, 
the following snapshots are commonly used: (I) $S=1, Y_L=0.4$; (II) $S=2, Y_{\nu}=0$; (III) $S=0, Y_{\nu}=0$, 
similar to those reported in~\cite{Steiner2000,Shao2011,Menezes2015,Chu2016}.
Here $S=s/n_b$ represents the entropy per baryon, 
$Y_L=\frac{n_e+{n_{\nu_e}}+{n_\mu }+n_{\nu_\mu}}{n_b}$ is the number of leptons (including electrons, muons and neutrinos) per baryon with trapped neutrinos, 
and $Y_{\nu}$ is the number of neutrino per baryon, where $Y_{\nu}=0$ indicates neutrino free case.

\section{Results}
\label{sec:results}
In this section, we present numerical results for PNS matter and bulk properties of PNS under the assumptions of isothermy, isentropy and fixed lepton fractions.
In the isothermal condition, the $\beta$-equilibrium and charge neutrality are satisfied without neutrinos.
While in the isentropic condition, neutrino free ($Y_\nu=0$) or neutrino trapping ($Y_L=\textit{constant}$) are considered.
\subsection{isothermal PNS matter}
%The relation between $\gamma$ and the maximum mass of NS (different $T$, $S/A$, $S/A$ plus fixed $Y_L$, models)
%$n_b-{center}$
%%%%%%%%%%%%%%%
% EOS, stiffness, T
%%%%%%%%%%%%%%%
\begin{figure*}[htb]
\includegraphics[bb=20 20 680 580,width=0.46\textwidth]{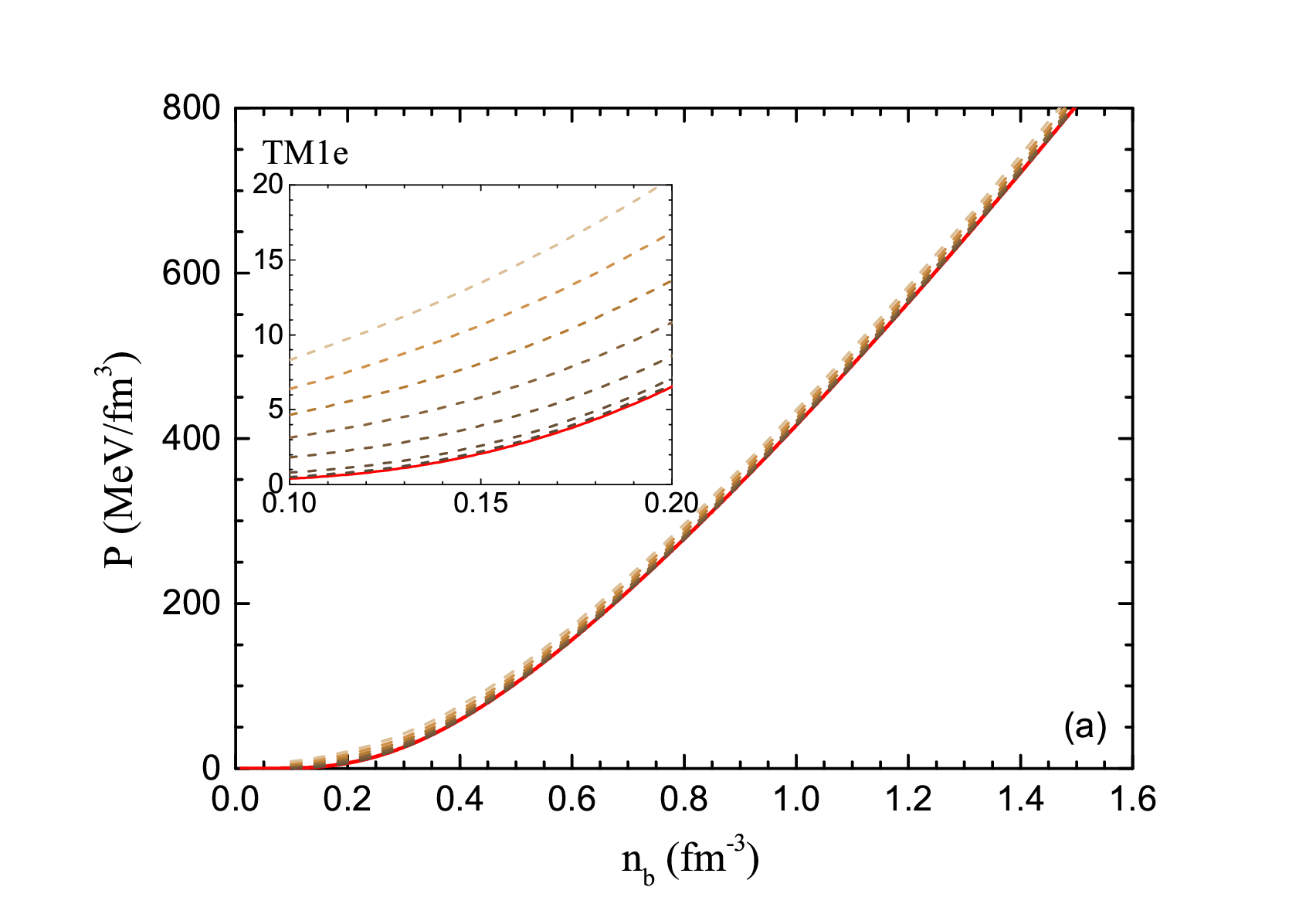} 
\includegraphics[bb=20 20 680 580,width=0.46\textwidth]{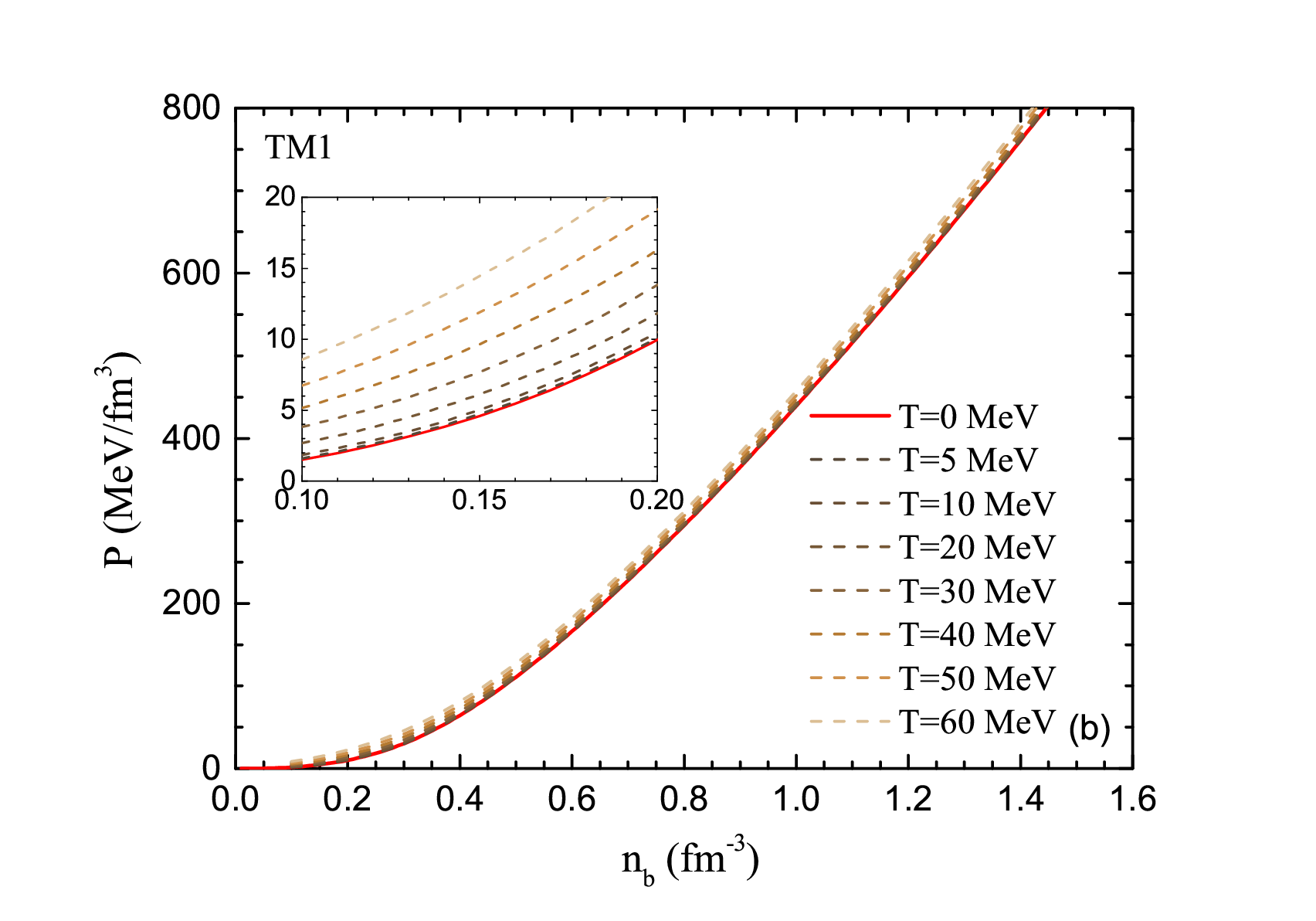} 
\caption{The  pressure $P$ as a function of baryon number density $n_b$ at different temperatures $T=0\sim60$ MeV under $\beta$-equlibrium.  
The TM1e results are presented in the left panel, and the TM1 results are in the right panel.
Sub-figures show an enlargement of the low-density region results.}
\label{fig:1nbp}
\end{figure*}
%%%%%%%%%%%%%%%

In Fig.~\ref{fig:1nbp}, we show the pressure $P$  as a function of baryon number density $n_b$ at different temperatures using TM1e and TM1 parameter sets.
It can be found that the thermal effect slightly increases the pressure $P$ by a few MeV across all densities, 
therefore, the thermal effect indeed decreases with density compared to the strong interaction among nucleons.
The differences between TM1e and TM1 results are also slight, 
with TM1 results being slightly stiffer at all temperatures $T$.
However, latter results will demonstrate that these differences have a significant impact on the bulk properties of PNSs.

%%%%%%%%%%%%%%%
\begin{figure*}[htb]
\includegraphics[bb=120 120 1080 980,width=0.95\textwidth]{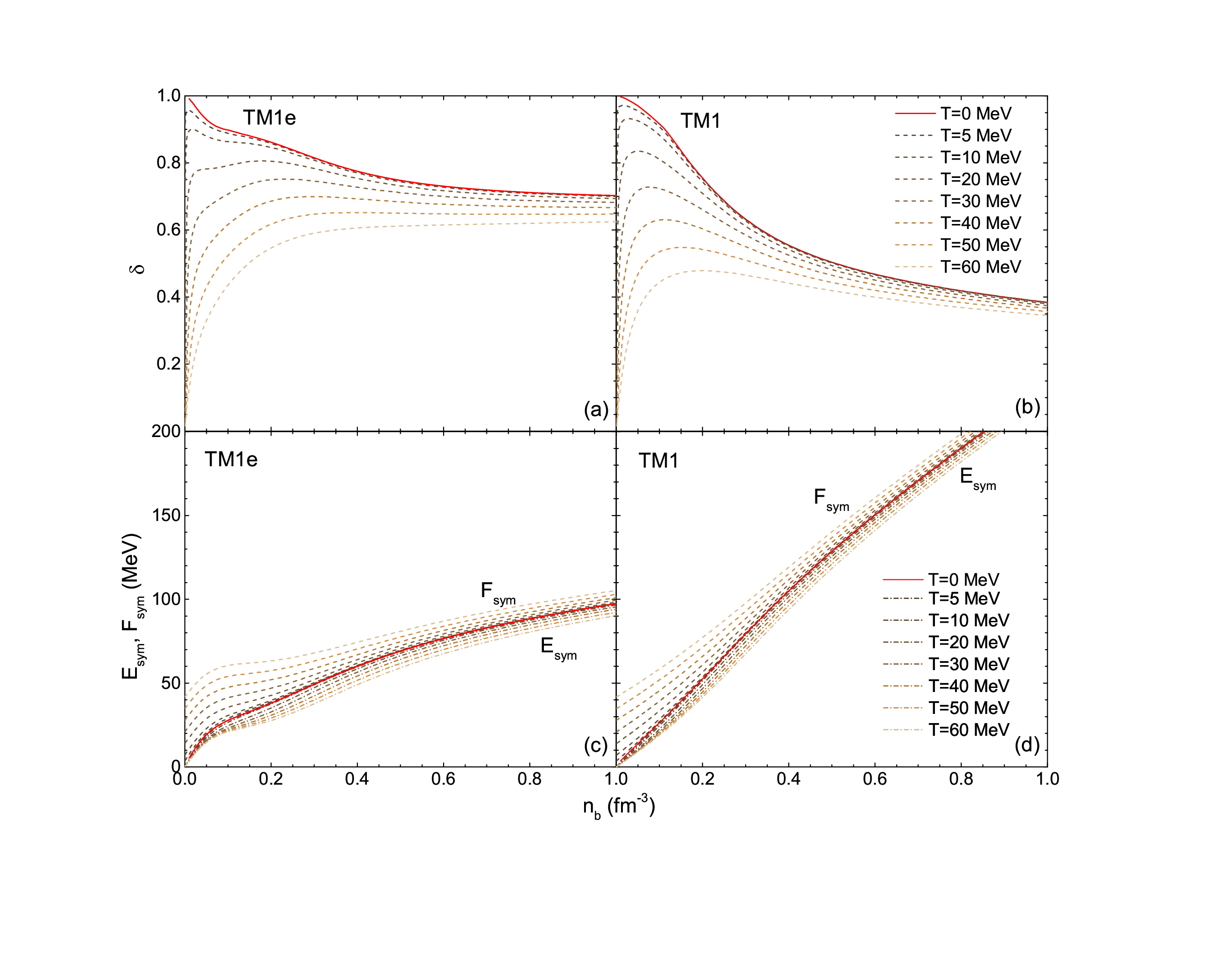} 
\caption{The isospin asymmetry $\delta$ (upper panel), the symmetry energy $E_{\textrm{sym}}$ (dash-dot lines, lower panel) and symmetry free energy $F_{\textrm{sym}}$ (dash lines, lower panel) as functions of the baryon number density $n_b$ under isothermal condition  ($T=0\sim60$ MeV). }
\label{fig:2nbesymfsym}
\end{figure*}
%%%%%%%%%%%%%%%%%%%%%%%%%%%%%%%

In Fig.~\ref{fig:2nbesymfsym}, we show the isospin asymmetry $\delta=\frac{n_n-n_p}{n_n+n_p}$ (upper panel) 
and the symmetry energy $E_{\textrm{sym}}$ and symmetry free energy $F_{\textrm{sym}}$ (lower panel) 
as functions of baryon number density $n_b$ for vary  temperatures $T=0\sim60$ MeV under isothermal condition.
$E_{\textrm{sym}}$ and $F_{\textrm{sym}}$ are calculated under the approximation
\begin{eqnarray}
F_{\textrm{sym}}&=&\frac{f}{n_b}(\delta=1)-\frac{f}{n_b}(\delta=0), \\
E_{\textrm{sym}}&=&\frac{\varepsilon}{n_b}(\delta=1)-\frac{\varepsilon}{n_b}(\delta=0).
\end{eqnarray}
Except for the differences arising from the symmetry energy slope $L$, 
the thermal effects for both TM1e and TM1 results are similar.
The isospin asymmetry $\delta$ is very sensitive to temperature $T$ at low densities, 
while the thermal effects becomes milder in the high-density region.
At lower densities (approximately $n_b<0.15$ fm$^{-3}$), a significant decline in $\delta$ with increasing $T$ indicates a transition to more symmetric matter.
There is a tendency for $\delta$ to converge at high densities, 
where higher densities lead to isospin-symmetric stable PNS matter.
In cold NS matter, the anti-$\beta$ decay, i.e. the neutralization procedure, can be explained as lowering the kinetic energy of electrons.
Therefore, with increasing temperature, higher thermal energy density and higher electron kinetic energy are expected.
This leads to an increase in electron number density, 
which in turn results in a higher proton number density to maintain charge equilibrium.
Consequently, the isospin asymmetry $\delta$ decrease with higher $T$.
Compared to TM1 (which has a higher slope $L$) results, $\delta$ curves for TM1e (with a lower slope $L$) level off more quickly. 
This difference is attributed to the slope $L$ of the symmetry energy.
TM1e has a slightly smaller symmetry energy $E_{\rm{sym}}$ at the saturation density, 
and the smaller $L$ causes $E_{\rm{sym}}$ to increase more gently compared to the TM1 results.
In the low-density region, the thermal effect has a greater impact on the symmetry free energy $F_{\textrm{sym}}$ than at high densities,
while the symmetry energy $E_{\textrm{sym}}$ changes more gradually  with different $T$.
The finite temperature effect causes $E_{\rm{sym}}$ to decrease 
while the symmetry free energy $F_{\rm{sym}}$ increases (nothing that $E_{\rm{sym}}=F_{\rm{sym}}$ when $T=0$),
which indicates that the impact from nucleon isospin on total energy density increases under finite temperature conditions.

%%%%%%%%%%%%%%%
\begin{figure*}[htb]
\includegraphics[bb=20 20 680 580,width=0.46\textwidth]{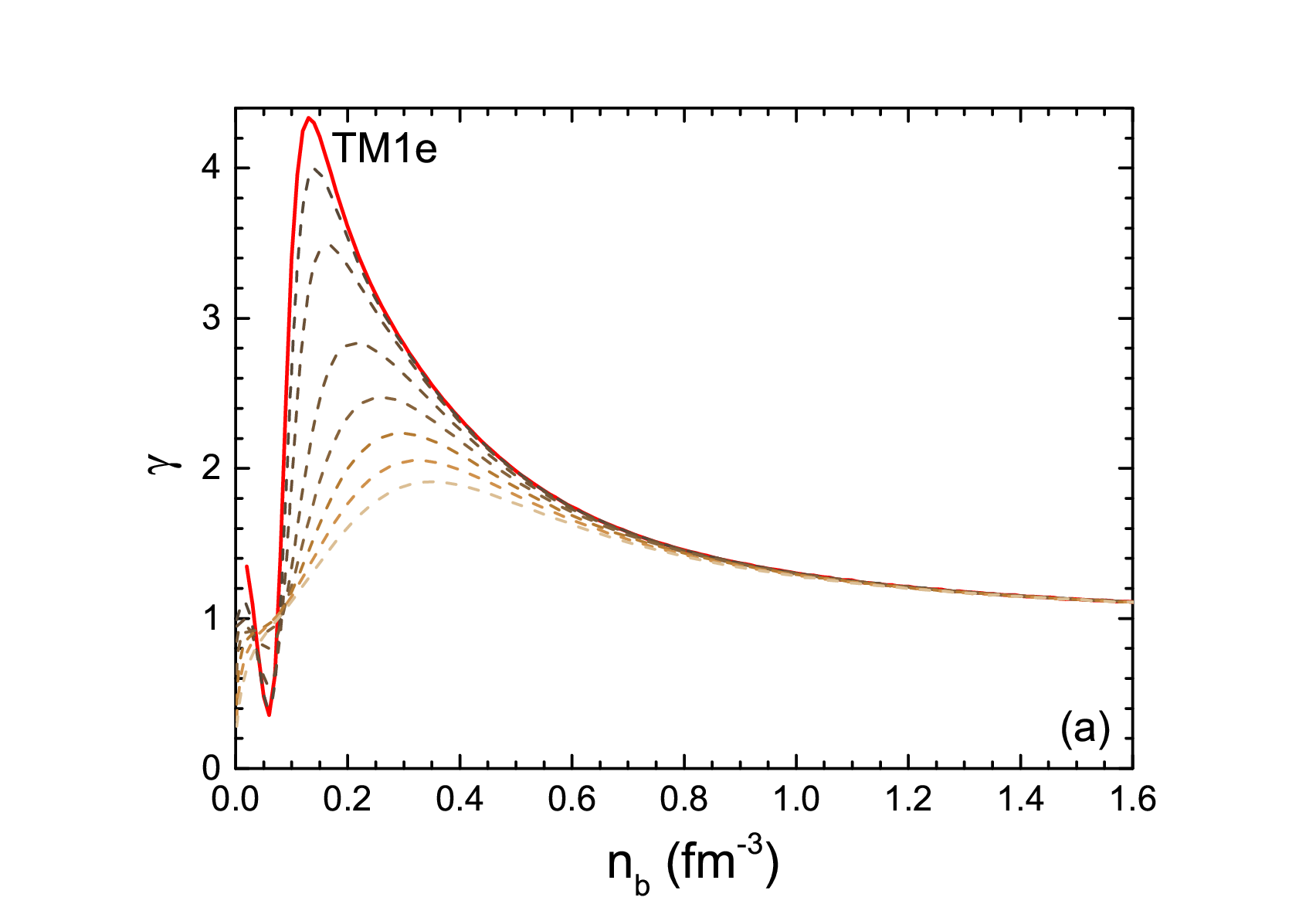} 
\includegraphics[bb=20 20 680 580,width=0.46\textwidth]{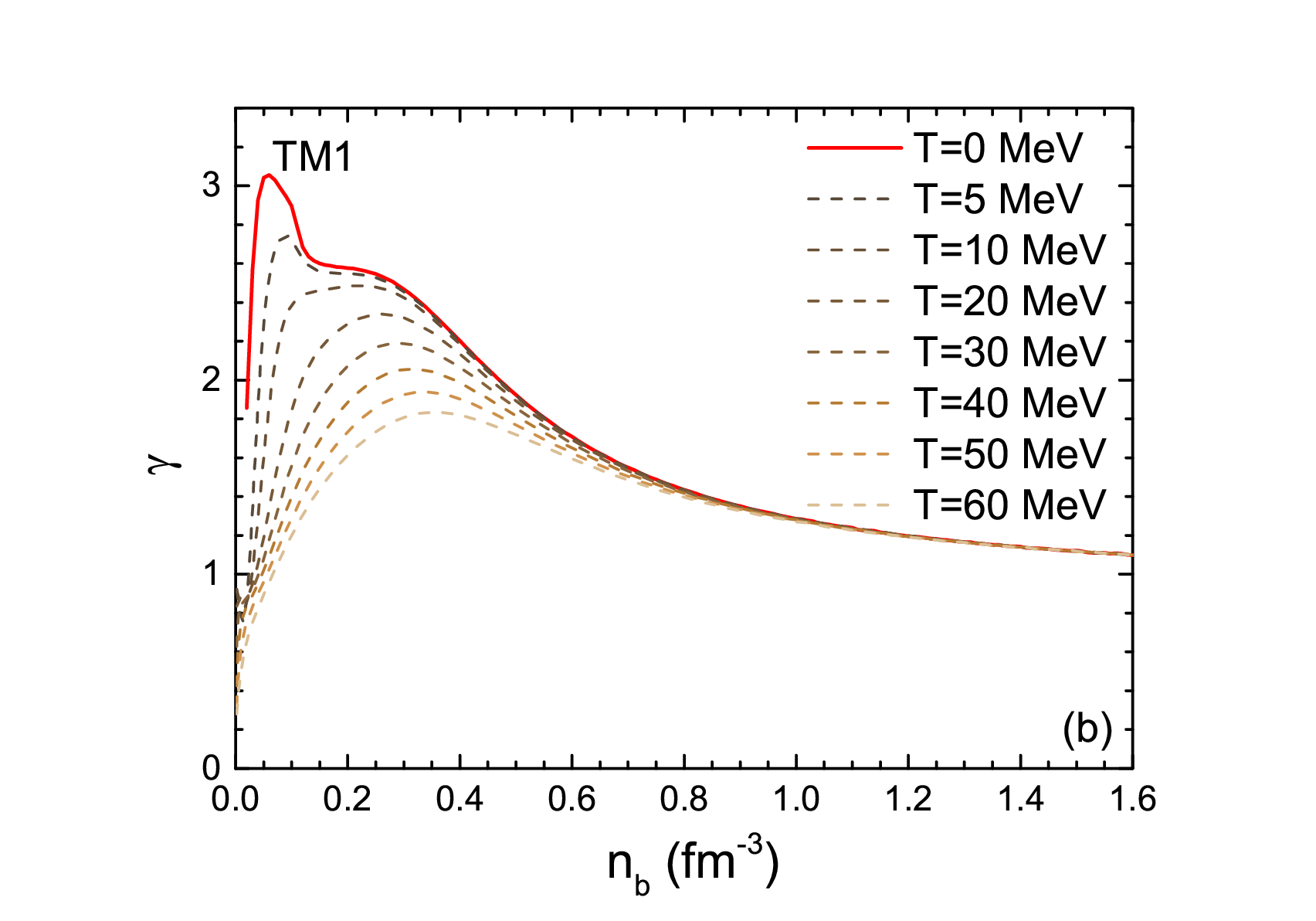} 
\caption{The polytropic index $\gamma$  as a function of the baryon number density $n_b$ at different temperature $T=0\sim60$ MeV. The results are derived from the EOSs in Fig.~\ref{fig:1nbp}}.
\label{fig:3nbgamma}
\end{figure*}
%%%%%%%%%%%%%%%%%%%%%%%%%%%%%%%
In Fig.~\ref{fig:3nbgamma}, we calculate the polytropic index $\gamma$ as a function of baryon number density $n_b$ at different temperatures,
with the expression given by 
$\gamma={\partial({\mathrm{ln}P})}/{\partial({\mathrm{ln}\varepsilon})}$.
It can be observed that with increasing temperature $T$, the curves of $\gamma$ decrease, 
showing a similar trend to the isospin asymmetry $\delta$.
As $T$ increases, the $\gamma$  curves become gentler, especially in the low-density region.
Despite the difference in symmetry energy slope between TM1e and TM1,
they both exhibit a decrease in the polytropic index $\gamma$ to approximately $\gamma\sim1.1$ in the high baryon density region.
This behavior arises because the isospin effect diminishes in the high-density regime  as $n_b$ increases.
Additionally, $\gamma$ decreases with temperature $T$ at the same $n_b$,
and the gap in $\gamma$ curves caused by different $T$ shrinks and approaches zero when $n_b>0.6$ fm$^{-3}$.
This change indicates a reduced thermal impact in the high-density region, 
which can be inferred from Fig.~\ref{fig:1nbp}, 
that the contribution of thermal effects to the total pressure diminishes as density increases,
as does the fraction of energy density contributed by thermal effects.
On the contrary, in low density regions, 
the thermal contribution to pressure and energy density fraction is relatively significant.
The similar shapes of the curves indicate that nucleon interactions play a decisive role in EOS, 
especially at high densities.
The thermal effect on the overall decline of $\gamma$ suggests that the relationship between pressure and energy density becomes smoother.
In Table.~\ref{tab:PNSpropertiesT-TM1e} and Table.~\ref{tab:PNSpropertiesT-TM1}, 
we summarize the isothermal bulk properties of PNSs using the TM1e and TM1 models at different temperatures $T$.
With the inclusion of thermal effects, both the maximum mass and radius increase with $T$, 
while the central density decreases.
We observe that there exists a minimum mass $M_{\rm{min}}$ for PNSs at high temperature $T$ (at lower $T$, the minimum masses are positioned at very high radius values that are not shown in Table.~\ref{tab:PNSpropertiesT-TM1e} and Table.~\ref{tab:PNSpropertiesT-TM1}), 
which has been discussed in Refs.~\cite{Haensel2002,Suwa2018,Koliogiannis2021}. 
$M_{\rm{min}}$ is the minimum value below which a non-equilibrium state is reached. 
It may provide insights into the minimum mass of the resulting remnant NS after CCSNe.
Our results show that the minimum mass $M_{\rm{min}}$ increases rapidly with temperature $T$,
such that for $T>30$ MeV, $M_{\rm{min}}$ may exceed the maximum mass of PNSs.
Beyond this temperature, the isothermal PNS is in a non-equilibrium configuration.
Combining Fig.~\ref{fig:3nbgamma} and Table.~\ref{tab:PNSpropertiesT-TM1e},~\ref{tab:PNSpropertiesT-TM1}, 
one can conclude that $\gamma$ has a negative correlation with the maximum star mass at finite temperatures within the TM1e or TM1 models.
The EOSs that can support larger maximum masses $M_{\rm{max}}$ are associated with lower $\gamma$ curves.
However, this relationship may not hold when considering different models, 
even though TM1e and TM1 differ only in their symmetry energy and its slope.

%%%%%%%%%%%%%%%%%%%%%
\begin{table}[htp]
	\caption{Summary of isothermal PNS bulk properties with neutrino free case for TM1e parameter set. 
 $M_{\rm max}$ is the maximum mass, $R_{\rm max}$ and $n_{b}^{c}$ are the corresponding radius and the central density. $R_{1.4}$ is the radius of the canonical 1.4~$M_\odot$ PNS. $M_{\rm min}$ and $R_{\rm min}$ are the mass and radius of the minimum mass PNS.}
 \begin{center}
		\setlength{\tabcolsep}{2.1mm}{
	\begin{tabular}{lcccccccccccc}
		\hline\hline
			 $T$  & $M_{\rm max}$ & $R_{\rm max}$ & $n_{b}^{c}$  & $R_{1.4}$ & $M_{\rm min}$  & $R_{\rm min}$\\
			 (MeV)  & $(M_{\odot})$ & (km) & ${\rm (fm^{-3})}$  & (km) & $(M_{\odot})$ & (km) \\
        \hline
              0   & 2.121 & 11.81 & 0.8959  & 13.02 & - & -\\
			 5   & 2.125 & 12.19 & 0.8953  & 14.02 & - & -\\
              10  & 2.129 & 12.85 & 0.8949  & 16.18 & - & -\\
              20  & 2.147 & 15.28 & 0.8708  & 30.08 & 1.265 & 47.21\\
			 30  & 2.192 & 22.10 & 0.8234  & - & 2.041 & 43.21\\
		\hline\hline
		\end{tabular}}
		\label{tab:PNSpropertiesT-TM1e}
	\end{center}
\end{table}

\begin{table}[htp]
	\caption{Same as Table~\ref{tab:PNSpropertiesT-TM1e}, but for TM1 parameter set.}
	\begin{center}
		\setlength{\tabcolsep}{2.1mm}{
	\begin{tabular}{lcccccccccccc}
		\hline\hline
			 $T$  & $M_{\rm max}$ & $R_{\rm max}$ & $n_{b}^{c}$  & $R_{1.4}$ & $M_{\rm min}$  & $R_{\rm min}$\\
			 (MeV)  & $(M_{\odot})$ & (km) & ${\rm (fm^{-3})}$  & (km) & $(M_{\odot})$ & (km) \\
        \hline
              0   & 2.178 & 12.37 & 0.8474  & 14.20 & - & -\\
			 5   & 2.180 & 12.72 & 0.8473  & 15.34 & - & -\\
              10  & 2.185 & 13.43 & 0.8468  & 18.02 & - & -\\
              20  & 2.206 & 15.88 & 0.8449  & - & 1.406 & 46.51\\
			 30  & 2.259 & 23.91 & 0.7414  & - & 2.162 & 42.72\\
		\hline\hline
		\end{tabular}}
		\label{tab:PNSpropertiesT-TM1}
	\end{center}
\end{table}

%%%%%%%%%%%%%%%
% stiffness, S, (S, YL)

%%%%%%%%%%%%%%%
\subsection{isentropic PNS bulk properties}
%%%%%%%%%%%%%%%%%%%%%
%%%%%%%%%%%%%%%
\begin{figure*}[htb]
\includegraphics[bb=120 10 1100 350,width=0.8\textwidth]{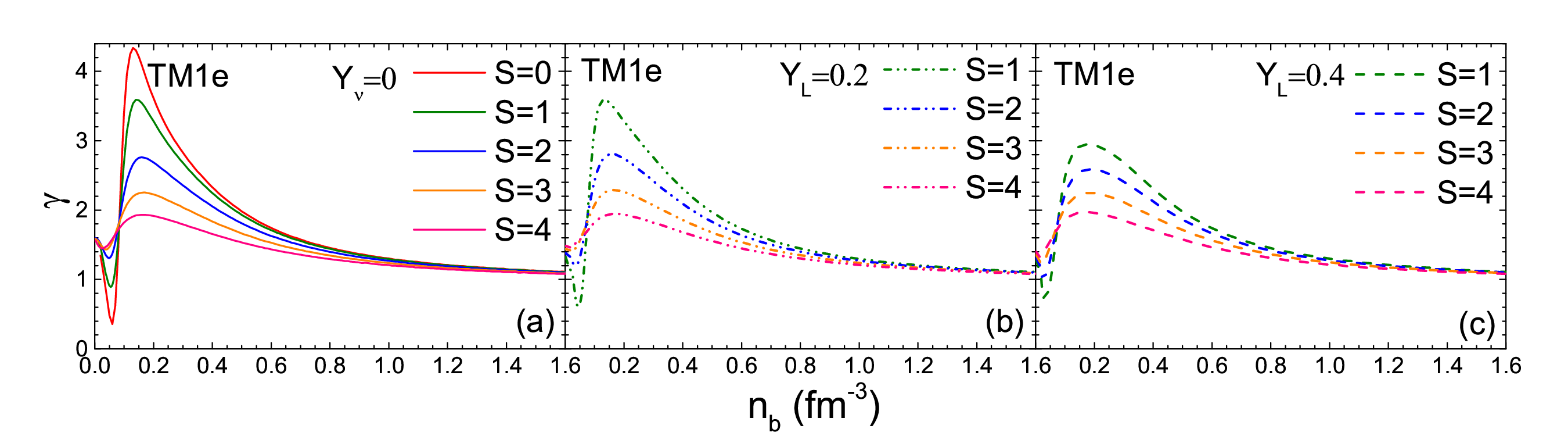} 
 \\
\includegraphics[bb=120 10 1100 350,width=0.8\textwidth]{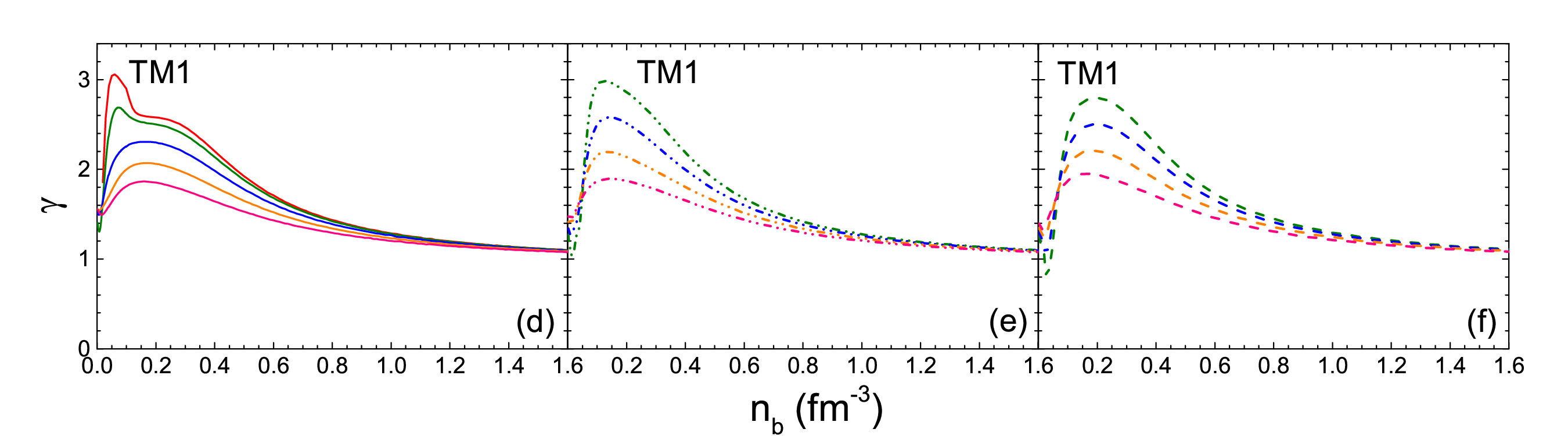} 
\caption{The polytropic index $\gamma$  as a function of the baryon number density $n_b$ at different entropy per baryon $S$, considering both neutrino free ($Y_\nu$=0, left panel) and neutrino trapping scenarios ($Y_L=0.2$ (middle pane), $Y_L=0.4$ (right panel)). }
\label{fig:4nbgammaS}
\end{figure*}
%%%%%%%%%%%%%%%%%%%%%%%%%%%%%%%

In Fig.~\ref{fig:4nbgammaS}, we present the polytropic index $\gamma$ as a function of baryon number density $n_b$ at different entropy per baryon $S$, considering both neutrino free ($Y_{\nu_e}=0$) and neutrino trapping ($Y_L=0.2, Y_L=0.4$) conditions.
As the $S$ increases, the curves of $\gamma$ become less steep, 
particularly at lower densities, exhibiting a trend similar to the isothermal curves shown in Fig.~\ref{fig:3nbgamma}.
This behavior can also be attributed to the thermal effects, under isentropic conditions, 
which lead to smoother changes in pressure relative to energy density.

A similar trend in the curves is observed at high densities.
However, in the low-density region, we find that $\gamma(Y_L=0.2)>\gamma(Y_L=0.4)>\gamma(Y_{\nu_e}=0)$ for the same entropy per baryon $S$.
The differences observed among $Y_{\nu}=0$, $Y_L=0.2$, and $Y_L=0.4$ results are affected by the contributions from neutrino to the pressure and energy density.
The distinctions between the neutrino-trapping cases
($Y_L=0.2$ and $Y_L=0.4$) stem from the neutrinos impact on thermodynamic quantities and the isospin differences.
The latter is expected to be the primary factor,
as $Y_L=0.4$ corresponds to more symmetric PNS matter , resulting in slightly softer EOSs.
In Table.~\ref{tab:PNSpropertiesS-TM1e} and Table.~\ref{tab:PNSpropertiesS-TM1}, 
we summarize the isentropic PNS bulk properties using TM1e and TM1 models under both neutrino free ($Y_{\nu_e}=0$) and neutrino trapping ($Y_L=0.2$ and $Y_L=0.4$) conditions.
As the entropy per baryon $S$ increases, a heavier PNS is obtained with a smaller central density $n_b^c$ but a higher central temperature $T_c$.
A lower value of $L$ (TM1e model) results in a higher central temperature $T_c$ for the same $S$ and $Y_L$.
Also, in conjunction with Fig.~\ref{fig:4nbgammaS}, one can conclude that $\gamma$ exhibits a negative relationship with the increase in PNS maximum mass under isentropic conditions.
This relationship is specific to the same model (TM1e or TM1) and the same isentropic condition (either $Y_{\nu_e}=0$ or same $Y_L$).

%%%%%%%%%%%%%%%
% PNS propertis 
%%%%%%%%%%%%%%%
\begin{table}[htp]
	\caption{Summary of isoentropic PNS bulk properties with neutrino free ($Y_{\nu}=0$) and neutrino trapping ($Y_L=0.2, 0.4$) for TM1e parameter set.
 $T_c$ is the central temperature of the maximum mass PNS.}
 \begin{center}
		\setlength{\tabcolsep}{2.1mm}{
	\begin{tabular}{lcccccccccccc}
		\hline\hline
		  $Y_L$ & $S$  & $M_{\rm max}$ & $R_{\rm max}$ & $n_{b}^{c}$ & $T_{c}~\text{}~\text{ }$ & $R_{1.4}$ \\
			& $(k_{B})$  & $(M_{\odot})$ & (km) & ${\rm (fm^{-3})}$ & (MeV) & (km) \\
   \hline
                  & 0  & 2.121 & 11,81 & 0.896 & 0 & 13.02 \\
		$\nu$   & 1  & 2.140 & 12.15 & 0.864 & 36.24  & 13.67 \\
		free   & 2  & 2.194 & 13.13 & 0.785 & 69.20  & 15.93 \\
		         & 3  & 2.296 & 15.08 & 0.660 & 93.94  & 21.89 \\
        \hline
		    & 1  & 2.120 & 11.92 & 0.922 & 35.30  & 13.57 \\
		0.2	& 2  & 2.179 & 12.83 & 0.839 & 68.63  & 15.60 \\
		    & 3  & 2.281 & 14.64 & 0.703 & 91.72  & 20.84 \\
        \hline
		    & 1  & 2.107 & 12.32 & 0.903 & 28.83  & 14.95 \\
		0.4	& 2  & 2.157 & 13.15 & 0.826 & 57.40  & 17.07 \\
		    & 3  & 2.230 & 14.73 & 0.705 & 82.70  & 21.97 \\
		\hline\hline
		\end{tabular}}
		\label{tab:PNSpropertiesS-TM1e}
	\end{center}
\end{table}

\begin{table}[htp]
	\caption{Same as Table~\ref{tab:PNSpropertiesS-TM1e}, but for TM1 parameter set.}
	\begin{center}
		\setlength{\tabcolsep}{2.1mm}{
	\begin{tabular}{lcccccccccccc}
		\hline\hline
		  $Y_L$ & $S$  & $M_{\rm max}$ & $R_{\rm max}$ & $n_{b}^{c}$ & $T_{c}~\text{}~\text{ }$ & $R_{1.4}$ \\
			& $(k_{B})$  & $(M_{\odot})$ & (km) & ${\rm (fm^{-3})}$ & (MeV) & (km) \\
   \hline
              & 0  & 2.178 & 12.37 & 0.847 & 0 & 14.20 \\
		$\nu$   & 1  & 2.193 & 12.58 & 0.841 & 31.14  & 14.78 \\
		free   & 2  & 2.242 & 13.52 & 0.760 & 61.17  & 17.08 \\
		   & 3  & 2.339 & 15.40 & 0.642 & 86.89  & 23.05 \\
        \hline
		  & 1  & 2.216 & 12.52 & 0.822 & 31.57  & 14.40 \\
		0.2	& 2  & 2.265 & 13.46 & 0.745 & 60.50  & 16.53 \\
		  & 3  & 2.357 & 15.18 & 0.651 & 84.10  & 21.85 \\
        \hline
		  & 1  & 2.107 & 12.35 & 0.893 & 29.04  & 14.95 \\
		0.4	& 2  & 2.157 & 13.23 & 0.806 & 57.30  & 17.07 \\
		  & 3  & 2.253 & 14.84 & 0.698 & 82.24  & 22.43 \\
		\hline\hline
		\end{tabular}}
		\label{tab:PNSpropertiesS-TM1}
	\end{center}
\end{table}

%%%%%%%%%%%%%%%
\begin{figure*}[htb]
\includegraphics[bb=5 20 680 580,width=0.46\textwidth]{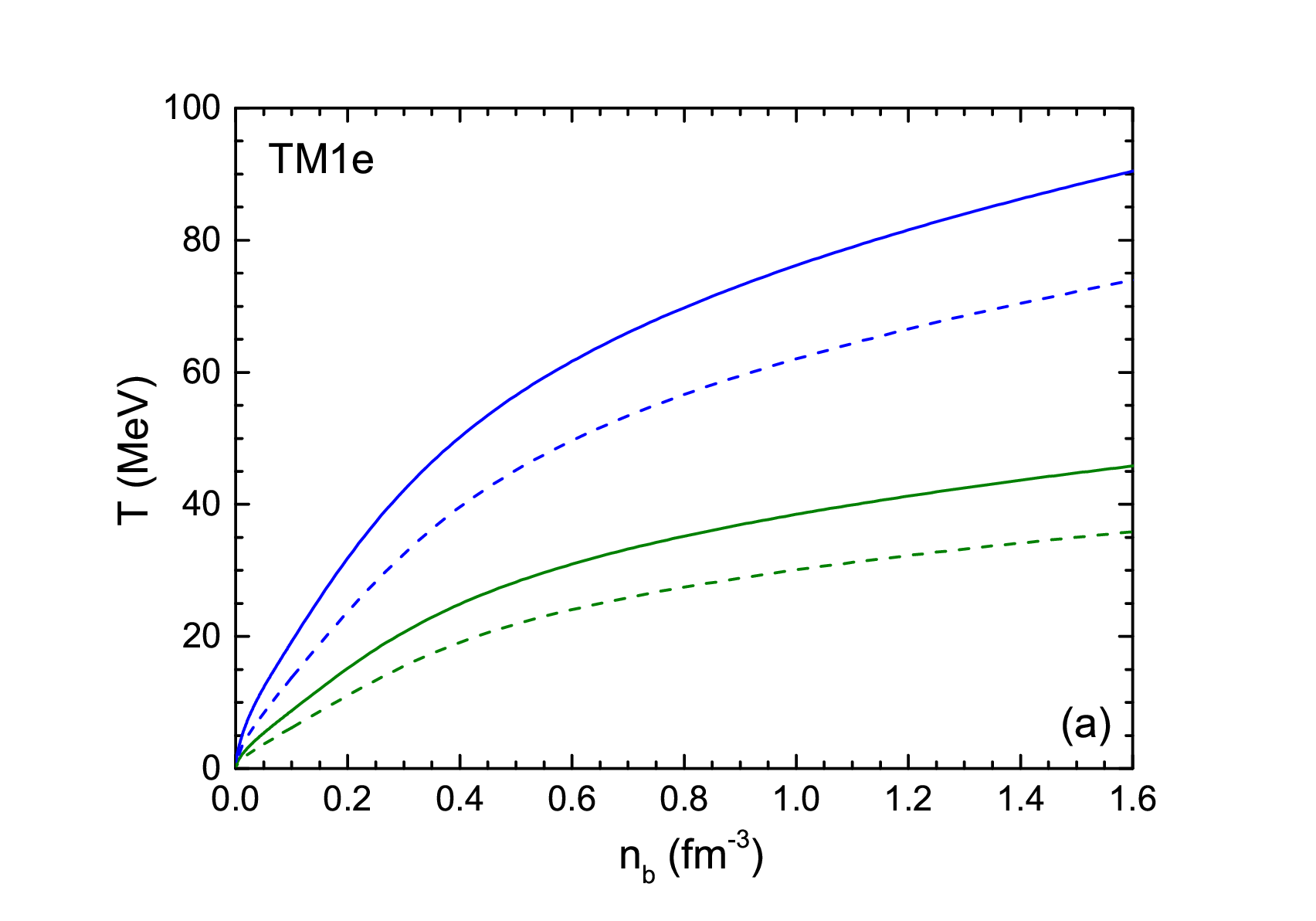} 
\includegraphics[bb=5 20 680 580,width=0.46\textwidth]{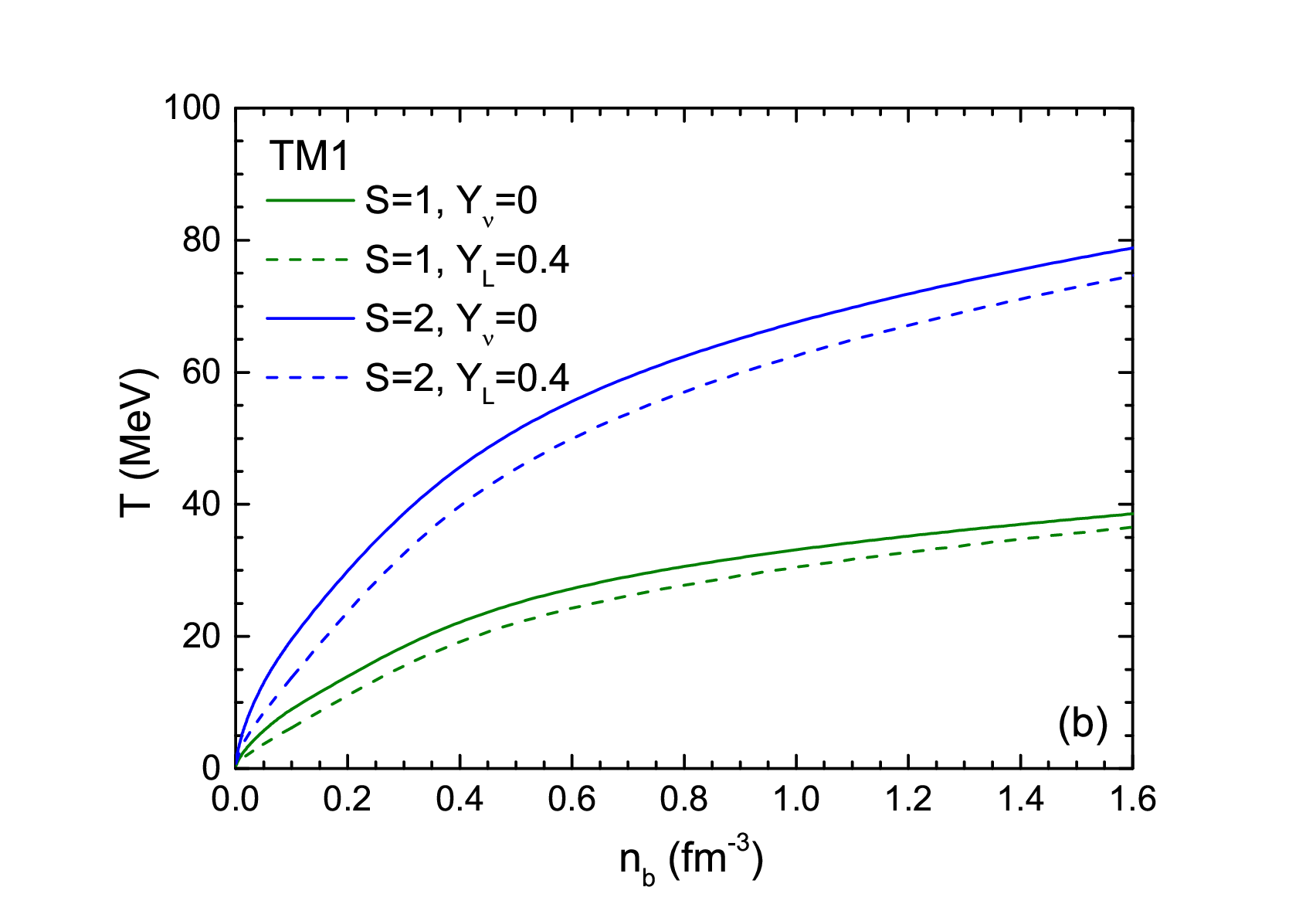} 
\caption{The temperature $T$ as functions of the baryon number density $n_b$. }
\label{fig:5nbt}
\end{figure*}
%%%%%%%%%%%%%%%%%%%%%%%%%%%%%%%
\begin{figure*}[htb]
\includegraphics[bb=15 25 680 580,width=0.46\textwidth]{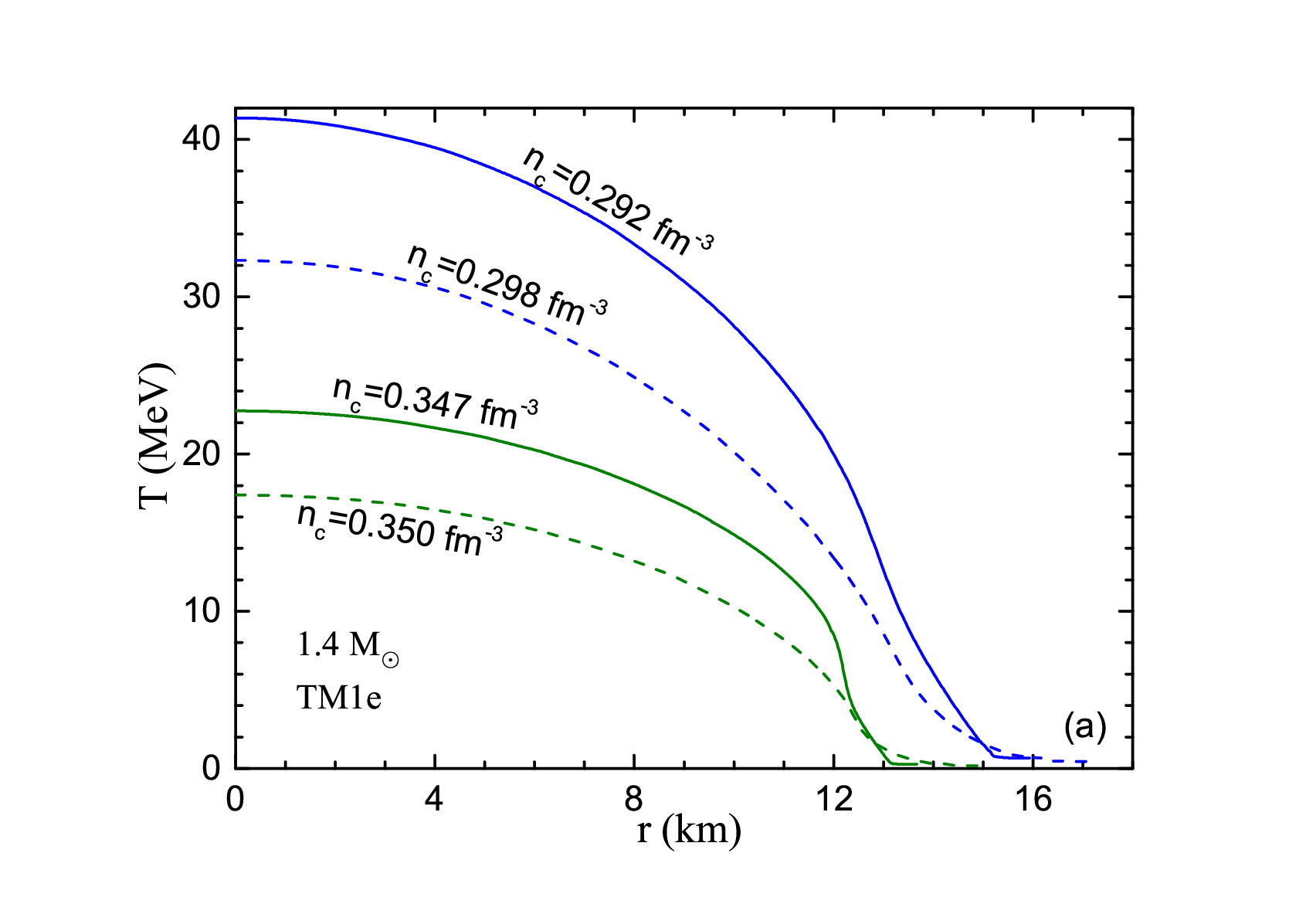} 
\includegraphics[bb=15 25 680 580,width=0.46\textwidth]{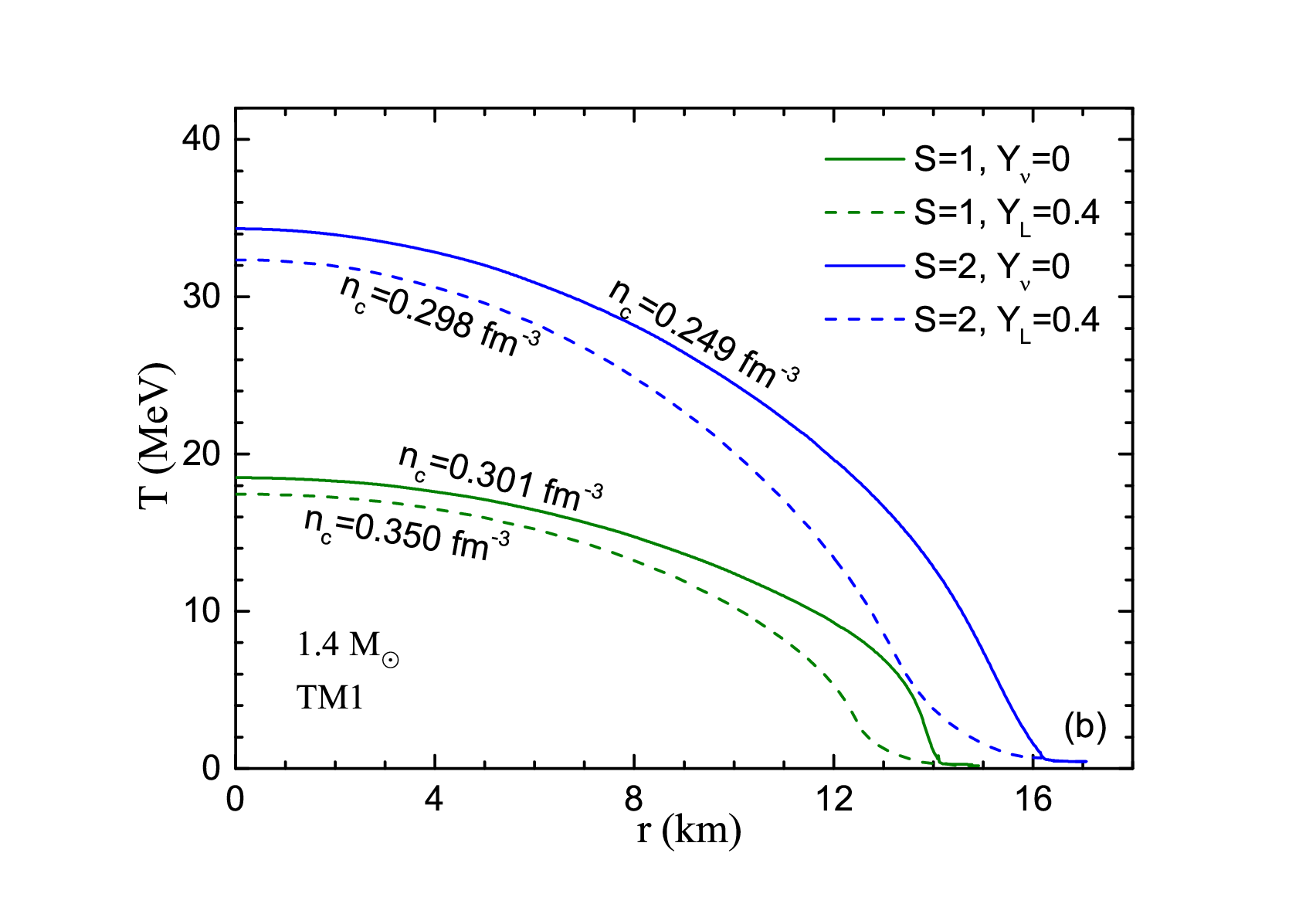} 
\caption{The temperature $T$ as functions of inner radius ($r$) in the canonical $1.4~M_\odot$ stars. The total radius of these stars $R_{1.4}$ can be found in Table~\ref{tab:PNSpropertiesS-TM1e} and \ref{tab:PNSpropertiesS-TM1}. $n_c$ are the central density correspond to the respective stars.} 
\label{fig:6RT-pns}
\end{figure*}
%%%%%%%%%%%%%%%%%%%%%%%%%%%%%%%

In Fig.~\ref{fig:5nbt}, we show the temperature $T$ as functions of $n_b$ for $Y_\nu=0$ and $Y_L=0.4$ at $S=1$ and $S=2$.
It can be seen the values of $T$ significantly decrease with decreasing density, 
especially near the PNS surface, where very low temperatures are evident.
There are notable differences between the TM1e and TM1 results,
specifically, TM1e exhibits a steeper increasing slope of $T$ with respect $n_b$.
The effect of $Y_L$ under the same entropy per baryon $S$ is  more pronounced for TM1e results than for TM1, 
indicating a larger influence from the isentropic condition due to the smaller symmetry energy slope $L$.
However, with higher symmetry ($Y_L=0.4$), 
the differences between TM1e results and TM1 results become less significant.
This is easily understood, as for symmetric nuclear matter ($Y_L=0.5$), TM1e and TM1 yield the same EOS. 
Since TM1 EOSs are slightly stiffer than TM1e (see Fig.~\ref{fig:1nbp}), we can infer under the isentropic situation, stiffer EOS are associated with lower temperatures with the same $n_b$.

In Fig.~\ref{fig:6RT-pns}, we show the temperature $T$ profile inside canonical 1.4 $M_\odot$ PNSs at the same stages as in Fig.~\ref{fig:5nbt}. 
We use the symbol $r$ to indicate the inner radius, distinguishing it from the total radius $R$.
The overall profile of $T$ resembles the $T-n_b$ relations in Fig.~\ref{fig:5nbt}, since the canonical 1.4 $M_\odot$ PNSs have similar central densities ($n_c\sim 0.29-0.35~\fm^{-3}$).
With higher symmetry ($Y_L=0.4$), 
TM1e and TM1 results have the same central density ($n_c=0.298~\fm^{-3}$ for $S=1$; $n_c=0.35~\fm^{-3}$ for $S=2$) and very similar $T$ profile as also shown in Fig.~\ref{fig:5nbt}.
While for the neutrino free situation, TM1e leads to higher inner temperatures but an earlier temperature decrease when approaching the star's crust. 
This is consistent with the result that TM1 results in a thinner crust compared to TM1e~\cite{Bao2015} due to $L$ difference.

%%%%%%%%%%%%%%%
\begin{figure*}[htb]
\includegraphics[bb=15 20 560 580,width=0.46\textwidth]{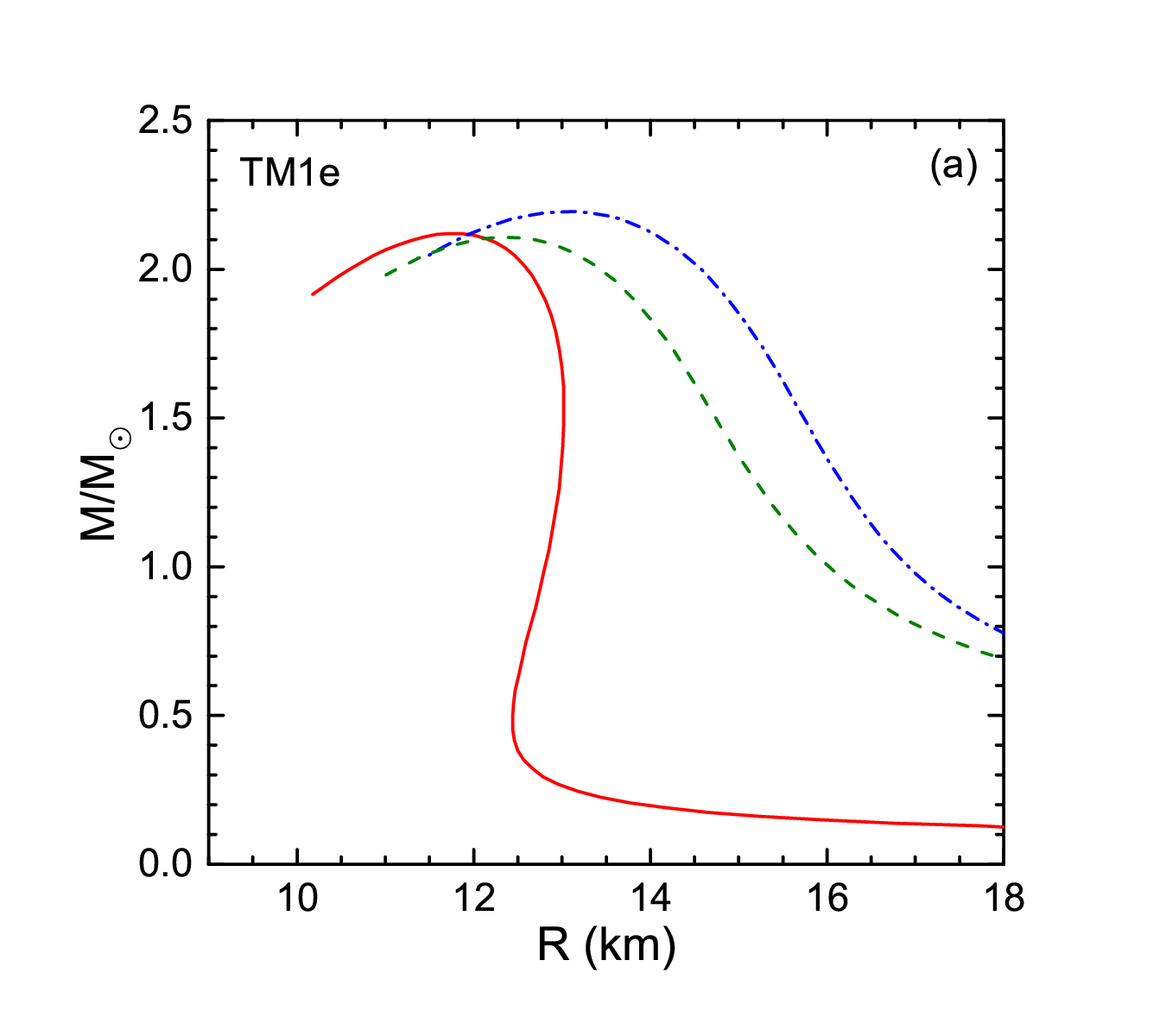} 
\includegraphics[bb=15 20 560 580,width=0.46\textwidth]{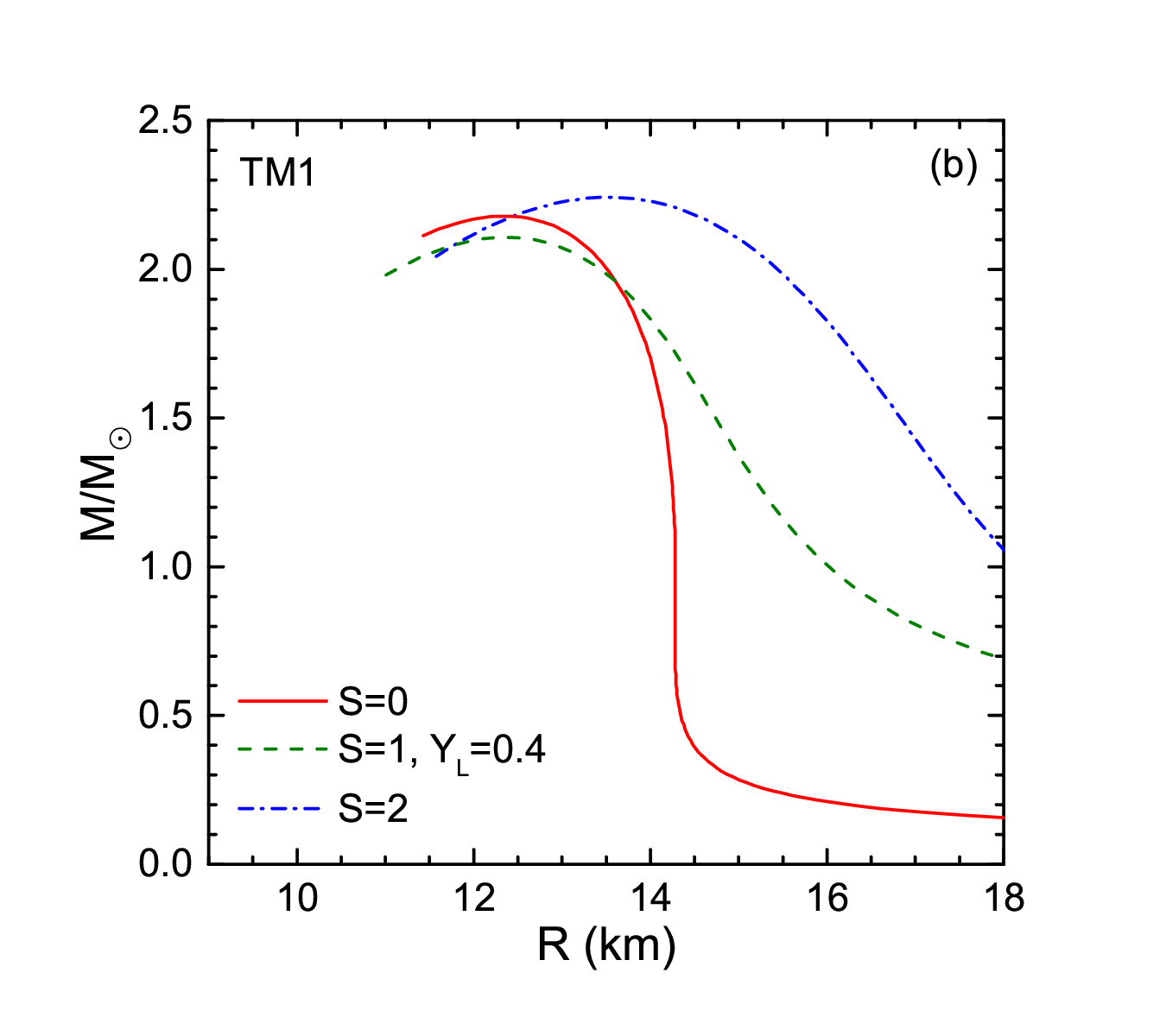} 
\caption{Mass-radius relation at different isentropic stage along PNS evolution line. }
\label{fig:7mr-pns}
\end{figure*}
%%%%%%%%%%%%%%%%%%%%%%%%%%%%%%%
Fig.~\ref{fig:7mr-pns} illustrates the mass-radius relations of PNSs at three isentropic stages along a PNS evolution line, 
utilizing both the TM1e and TM1 models.
Initially, during the heating stage, the PNS expands with increasing entropy per baryon $S$ 
while experiencing a decrease in the lepton fraction $Y_L$, i.e. the neutronization process.
As trapped neutrinos begin to escape, the star begins cooling.
In the first stage of PNS evolution, 
with an entropy per baryon $S=1$ and $Y_L=0.4$, 
there is a significant presence of trapped neutrinos, resulting in a maximum mass of 2.107~$M_\odot$ applying TM1e model, 
which is slightly smaller than the cold NS maximum mass of 2.121$M_\odot$.
On one hand, pure thermal effect leads the stellar expand thereby allowing for a larger maximum mass and radius, shown by the $S=2$ and $S=0$ curves.
We can also found in Table.~\ref{tab:PNSpropertiesS-TM1e} and Table.~\ref{tab:PNSpropertiesS-TM1} that in the neutrino-free case, the maximum mass increases with $S$.
On the other hand, due to the presence of trapped neutrinos,
PNS matter exhibits greater symmetry compared to cold NS matter. 
More symmetric nuclear matter results in relatively lower nucleon interactions compared to asymmetric nuclear matter, 
consequently yielding a lower maximum mass.
This relationship between the maximum masses in neutrino-free and neutrino-trapping cases is further illustrated in Table.~\ref{tab:PNSpropertiesS-TM1e} and Table.~\ref{tab:PNSpropertiesS-TM1}, where with a same $S$, 
a higher lepton fraction $Y_L=0.4$ (indicating more symmetric PNS matter) leads to a slightly diminished maximum mass compared to $Y_L=0.2$ or $Y_\nu=0$.
Comparing TM1 results with TM1e, the isentropic stage curves shift more significantly than those of the cold NS mass-radius relation.
Aside from differences in symmetry energy (and its slope), TM1e predicts higher temperatures at the same density for equal $S$ and $Y_L$, as shown in Fig.~\ref{fig:5nbt}.
Notably, the temperature differences also stem from the variations in symmetry energy and its slope, as discussed in Fig.~\ref{fig:5nbt}.

\section{Summary}
\label{sec:summary}

In this work, we calculate effect of symmetry energy on the EOS of hot nuclear matter and the bulk properties of PNSs.
We employ the RMF framework and introduced two parameter sets, TM1e and TM1, which preserve the same saturation properties except differing only in symmetry energy and its slope. 
To investigate the thermal effect, we consider conditions of isothermy, isentropy, and neutrino trapping with fixed lepton fractions.

In the isothermal case, thermal effects directly contribute to the pressure 
but reduce both isospin asymmetry and symmetry energy.
We find that the influence of thermal effects diminishes at high densities, 
where nuclear interactions become more significant.
In the isentropic case, neutrinos contribute less to the thermodynamic quantities compared to electrons and muons.
Additionally, a larger lepton fraction $Y_L$ indicates a more symmetric nuclear matter environment, 
resulting in lower corresponding energy and pressure compared to neutrino free nuclear matter.
Under the isentropic condition, a rapid temperature decrease is observed near the crust of PNS.
Lower $L$ (TM1e) leads to higher inner temperatures with neutrino free.

We calculate the bulk properties of PNSs at finite temperature and different snapshot stages along the evolution line. 
Our findings indicate that the central density decreases with increasing temperature or entropy.
Additionally, we observe that the polytropic index curve of EOS exhibits a negative correlation with the maximum mass of the PNS within the same model, regardless of whether the case is isothermy, isentropy, neutrino free or neutrino trapping.
%Usually a stiffer EOS could support a higher maximum mass of PNS, 
%therefore we believe the speed of sound $c_s^2$ rather than the polytropic index $\gamma$ could represent the EOS stiffness. 
%Compared with the speed of sound $c_s^2$, the polytropic index $\gamma$ more likely represent the deviation from free fermi gas, which has $\gamma=1$

Overall, our results indicate that symmetry energy effect significantly influence the EOS, the isospin asymmetry, the polytropic index
as well as the properties of PNSs at finite temperatures and evolution stages.
Especially during the heating stage, 
the mass-radius relations differ considerably from those of cold NSs due to the symmetry energy effect.

%====================================================================================================
%\nolinenumbers
\section*{Acknowledgements}
This work is supported by the National Natural Science
Foundation of China under Grants No. 12305148, No. 11975132, No. 12205158,
Hebei Natural Science Foundation No. A2023203055,
and the Shandong Provincial Natural Science Foundation, China under Grants No. ZR2023QA112, No. ZR2022JQ04, 
No. ZR2021QA037, and No. ZR2019YQ01.

\end{document}